\documentclass[12pt]{article}
\usepackage{amsmath,amssymb,pgf,pgfarrows,pgfnodes,subfig,float,appendix, hyperref}
\usepackage[margin=0.7in]{geometry}

\title{{\rm\footnotesize \qquad \qquad \qquad \qquad \qquad \ \qquad \qquad \qquad \ \ \ \ \ \                    RUNHETC-2013-2     
SCIPP 13/12}\vskip.5in     Dirac Gluinos in the Pyramid Scheme}
\author{Tom Banks\\
Department of Physics and SCIPP\\
University of California, Santa Cruz, CA 95064\\
{\it and}\\
Department of Physics and NHETC\\
Rutgers University, Piscataway, NJ 08854\\
E-mail: \href{mailto:banks@scipp.ucsc.edu}{banks@scipp.ucsc.edu}}
\begin{document}
\maketitle

\begin{abstract}
I point out several terms in the low energy effective Lagrangian of the Pyramid Scheme, which were missed in a previous analysis of the phenomenological consequences of the model.  They lead to a Dirac contribution to the gluino mass, much larger than the one loop Majorana mass.  The gluino can thus be much heavier than in previous estimates, without introducing corresponding large loop corrections to squark masses.  As pointed out by a number of authors,
this ameliorates the tension between the predictions of the model, and LHC data.  I also point out that the model has corrections to the Higgs potential, both at the tree and loop levels, which may ameliorate fine tuning.
\end{abstract}
\section{Introduction}

The theory of cosmological SUSY breaking\cite{CSB} says that the terms in the low energy effective Lagrangian, which violate a discrete R symmetry, and lead to spontaneous breaking of SUSY, come from diagrams where a gravitino propagates out to the horizon of de Sitter space and interacts with the vast set of degrees of freedom there.  This is only self consistent, if the gravitino mass is related to the cosmological constant by  $$m_{3/2} = K \Lambda^{1/4} .$$ Estimates based on the idea\cite{witten} that the ratio between the unification scale and the Planck scale is related to the volume of extra dimensions, which is also the unification scale, give $K \sim 10$.  The implied scale of splitting in non-gravitational supermultiplets is
$$F \sim 20-30 ({\rm TeV})^2 .$$  

One of the basic principles of Holographic Space-Time (HST), the theoretical framework underlying the hypothesis of CSB,  is that effective quantum field theory (QUEFT) is only a good approximation to processes in a given causal diamond in $4$ dimensions, when the total entropy accessed in the process is bounded by $A^{3/4}$, where $A$ is the area, in Planck units of the holographic screen of the diamond.  In other processes, very low energy horizon degrees of freedom, which are not contained in QUEFT, are important to a correct description of the physics.  For the maximal causal diamond in Minkowski space,
the horizon DOF decouple, and QUEFT is good to all orders in the expansion of scattering amplitudes in powers of kinematic invariants in Planck units, as long as the number of finite energy particles in the process is finite.  This means that a QUEFT description of these amplitudes is valid, with the same kinematic restrictions.

In dS space, there are interactions between particle DOF and the finite distance horizon, whose origin cannot be described in QUEFT.  However, for large radius, an approximate description in terms of Feynman diagrams is possible.  The important diagrams involve a virtual gravitino propagating out to the horizon and interacting with the Area entropy of DOF there.  The resulting amplitudes are of order $$ e^{-2 m_{3/2} R} \sum V^{\dagger} \frac{1}{\Delta E} V , $$ where $V$ is an operator describing emission and absorption of the gravitino by the horizon.  The density of states on the horizon is a few bits per unit Planck area, and the virtual gravitino can propagate on the horizon for a proper time of order $\frac{1}{m_{3/2}} $, because it is a massive particle and the horizon is null.  The convention random walk formula for the gravitino propagator, with a Planck scale proper time cutoff, implies that the area covered by the propagating gravitino is $\frac{1}{m_{3/2} M_P} $.  This implies that the gravitino-horizon graphs contribute to localizable low energy interactions of magnitude
$$e^{ - 2 m_{3/2} R + c \frac{M_P}{m_{3/2} }},$$ to exponential accuracy in the dS radius as $RM_P \rightarrow \infty$.  $c$ is a positive constant. In particular, the gravitino mass itself should be proportional to such a factor.  On the other hand, the restoration of SUSY in the Minkowski limit, implies that the gravitino mass should go to zero.  However, the above formula either blows up or goes to zero exponentially unless $$ 2 m_{3/2}^2 = c \frac{M_P}{R} ,$$ and this gives $$m_{3/2} = K \Lambda^{1/4} \sim 10^{-3} K\ {\rm eV}. $$   More refined considerations suggest that $K \sim 10$ if the size of compact dimensions is $ (2 \times 10^{16}\ {\rm GeV})^{-1} $.   Furthermore, if we assume $m_{3/2}$ goes to zero more rapidly than this formula, then our estimate for it blows up exponentially, while if we assume it goes to zero less rapidly, then our estimate goes to zero exponentially.  Thus,  $m_{3/2} \sim \Lambda^{1/4}$ is the only self consistent scaling law for the gravitino mass, if the origin of SUSY breaking comes from interaction with horizon DOF.  

The Pyramid Schemes\cite{pyramid} (reviewed below) are at present the only viable effective field theory descriptions, which both incorporate these ideas, and are consistent with gauge coupling unification and the absence of superpartners or other exotica in all extant experiments.  It is hard to make very precise predictions about the particle spectrum in the Pyramid models, because they contain a new strongly coupled sector with a confinement scale in the TeV range.  The Kahler potential on the moduli space of this strongly coupled SUSic gauge theory, enters into the potential for the Higgs sector, as well as the expressions for super-partner masses. In \cite{tbtj} we made a crude estimate of the spectrum, based on an uncontrolled approximation to the Kahler potential.  We were able to fit all LHC discoveries and bounds, with a fine tuning of a few percent.   

Recently, I realized that we had left off two important terms in the effective Lagrangian for the Pyramid models.  They are D-terms of fairly high minimal dimension, but if the ratio between the SUSY breaking scale $F_M$ and the confinement scale $\Lambda_3$ of the new strongly coupled $SU_P (3)$ gauge group at the apex of the Pyramid, is slightly greater than $1$, they make important modifications to the spectrum.   In this paper I will show that they lead to a Dirac mass for the gluino, coupling it to a composite chiral adjoint.  The Majorana mass of the adjoint is naturally of the same size as the Dirac mass, as a consequence of the fact that the QCD coupling $g_3 = \sqrt{4\pi \alpha_3}$ at the TeV scale is $\sim 1$.   Thus we get two Majorana gluinos, of comparable mass, which are roughly equal mixtures of the original gluino field and the fermion from the chiral adjoint. The leading contribution to the squared squark masses comes from a one loop QCD diagram with this massive Dirac gluino in the loop.  The model thus predicts $m_{\tilde{q}} \sim \sqrt{\frac{\alpha_3}{2\pi}} m_{1/2}^{(3)}$.   The correct interpretation of LHC bounds in this heavy gluino scenario is thus that the squarks are heavier than $800-900$ GeV and the gluino weighs $ 8 - 9$ TeV\footnote{There are strong interaction uncertainties in the gluino estimate.  It could easily be a factor of two heavier or lighter.} .

I also point out that for values of the parameters favored by the observed mass of the Higgs boson, there is potentially a large F term contribution to top squark masses, which will split half of them above, and half below the mass of the squarks of the first two generations.  

For the other gauginos, the Dirac mass is smaller by a factor of $g_1$ or $g_2$ (the standard model couplings) than the Majorana mass of the corresponding composite chiral adjoint, which is also (see below) larger than the Majorana mass of the gluino's partner. Thus, these particles get a Majorana seesaw mass, which is nominally a factor of $16 \pi^2$ larger than the normal gauge mediated contributions to their masses.  There is considerable strong interaction uncertainty in these estimates.

After a brief review of the Pyramid scheme, I will present the operators that give rise to the Dirac gluino scenario.  In the final section I'll outline the challenges involved in getting a decent estimate of the Higgs potential in these models.

\subsection{Review of the Pyramid Scheme}

The strategy for exploring the phenomenological implications of this mechanism is to write a low energy effective field theory describing the limiting model with $\Lambda = 0$.
This model must be exactly supersymmetric (because SUSY is a gauge symmetry and because the hypothesis of CSB is that it is not spontaneously broken in this limit) and preserve a discrete R-symmetry (to guarantee naturally vanishing c.c. in effective field theory).  The terms coming from horizon interactions cannot break SUSY explicitly, but they can break the discrete R symmetry.  Indeed, the gravitino mass breaks all R symmetries to $(- 1)^F$. To be consistent with an underlying theory of CSB, this must lead to {\it spontaneous} breakdown of SUSY.  The R-violating terms coming from the horizon interactions do not satisfy the field theory conditions of naturalness, because they come from very special diagrams.  In particular, the constant in the superpotential is chosen to tune the c.c. to the value consistent with
$$m_{3/2} = K \Lambda^{1/4}.$$  Further violation of naturalness allows us to evade the Nelson-Seiberg theorem\cite{nelsonseiberg} and write a non-generic superpotential, which spontaneously breaks SUSY in the absence of R-symmetry.

We choose the R-symmetry to obey certain phenomenological constraints.  It forbids all terms (for $\Lambda = 0$) of dimension $\leq 6$, which violate $B$ or $L$, except for the dimension $5$ operator that is responsible for neutrino masses.  Even in the presence of R violation, these terms do not reappear, since all R violating diagrams have a pair of gravitino lines emanating from the horizon, connected to a diagram in the $\Lambda = 0$ theory.  These terms still preserve, with the same exception,  $B$ and $L$ up to dimension $6$.  We also choose the R charges to be flavor blind, and to forbid the $\mu$ term $H_u H_d$.  The latter however may be generated via R violating diagrams, so it will scale to zero with a (perhaps fractional) power of the gravitino mass.  We assume it's no bigger that the TeV scale so that the Higgs fields remain in the effective Terascale model.  This is of course natural, since $\sqrt{F} \sim 5$ TeV.  

Significant constraints on the model come from the non-observation of gauginos.   With such a low SUSY breaking scale, we are forced to a model of direct mediation.   That is, there must be a new strongly coupled gauge sector, with group $G_P$ and confinement scale $\Lambda_3 \sim $ a few TeV, which couples directly to the SUSY breaking sector and contains particles charged under the standard model.  Coupling unification puts strong constraints on the new gauge group and matter content and we have found that the only plausible models are the Pyramid Schemes described below.   These utilize trinification\cite{trinification}, and contain new chiral fields (the {\it trianons}) $T_i \oplus \tilde{T}_i $, 
in the $$(F, 1, \bar{3}_i , 1) \oplus (\bar{F}, 1, 3_i , 1) $$ for each of the $SU(3)_i$ groups of the trinified standard model.  We of course assume that trinification is spontaneously broken at the unification scale and that only the MSSM fields, the trianon fields and some standard model singlets survive this breaking.  $F$ is the fundamental representation of the Pyramid group $G_P$.  $G_P$ is restricted by the requirement that standard model couplings remain perturbative up to the unification scale. It could be $SU_P (N)$ with $N = 3,4$, but the $N = 3$ model is much nicer in many ways, so we will assume it here.   Below the confinement scale $\Lambda_3$ the physics of this model is described by Seiberg's effective Lagrangian, which contains singlet and color octet meson fields, $M$ and $M^a$ which are components of the meson matrix $M_i^j = (T_3)^a_i (\tilde{T}_3)_a^j$.

As it stands, the model does not achieve the goal we set for it.  The only operator of dimension $\leq 4$, which can be chosen to violate R symmetry and be induced by interactions with the horizon is the $\mu$ term of the MSSM, and mass terms for trianons.  With or without these terms, the model does not violate SUSY.  The simplest way to remedy this, which we have adopted, is to add three new singlet fields $S_i$ which we couple to the trianons and Higgs fields via couplings in the superpotential
$$ W = \sum (\alpha_i^j S^i + m_j) T_j \tilde{T}_j  + (\beta_i S_i + \mu ) H_u H_d + C(S) 
+ \sum \lambda_j (T_j )^3 + \sum \tilde{\lambda}_j (\tilde{T}_j )^3 + W_{MSSM} + W_0.$$  We have constructed the discrete R symmetry so that the cubic terms in this super-potential have R charge $2$, while the lower order terms violate the R symmetry.  One of the $\lambda_j, \tilde{\lambda}_j$ pairs should vanish in order to have a simple dark matter candidate in the model. $C(S)$ is a cubic polynomial in the singlets.  It should be chosen such that SUSY is unbroken when the R symmetry is preserved and spontaneously violated when the R violating terms are taken into account.  A simple way to do this is to make all the linear and bilinear terms in the super-potential violate the R symmetry.  Then $S_i = H_u = H_d = M = 0$ is a supersymmetric solution.   A superpotential with R violating linear terms $F_i S_i$, with $F_i$ linearly independent of both $\alpha_i$ and $\beta_i$ and appropriate restrictions on the bilinear and trilinear terms, so that the determinant of $C_{ijk} S_k + C_{ij}$ vanishes for all $S_i$ guarantee SUSY violation.  These restrictions do not satisfy the genericity conditions of Nelson and Seiberg, but we have already said that these do not apply to the R-violating terms coming from the horizon.

If the Kahler potential were canonical, the SUSY violating $F$ component would decouple from the standard model fields.  We would obtain a model that implemented the ideas of CSB, but did not match phenomenology.  Fortunately, the Kahler potential has non-trivial dependence on two linear combinations of the $S_i$, which couple to the electroweak trianons.   In \cite{tbtj} we calculated the dependence of $K$ on $S_i$ assuming that the Pyramid gauge coupling was perturbative at the scale of the electro-weak trianon masses, and used this result to get an estimate for superpartner masses in the MSSM.   There is no reason for this to be a good approximation to the version of the model relevant to the real world, but it was the best we could do.

The $SU_P (3)$ gauge group is not asymptotically free above the scale of the trianon masses.   However, if all three couplings $\lambda_p$ and $\tilde{\lambda}_p$ were present there would be a line of fixed points in the model in which these couplings are set equal to the $SU_P (3)$ gauge coupling, which can take any value.   If that coupling is fairly strong at high energies, then as soon as two of the trianons decouple the model quickly becomes confining, and its behavior below the confinement scale is described by Seiberg's meson/baryon Lagrangian\cite{nati} for $N_F = N_C = 3$.  In \cite{bfk} we showed that behavior similar to this occurred even if one of the pyrma-baryon number violating trilinear coupling pairs vanished\footnote{We wanted one of these accidental symmetries to be preserved at the renormalizable level, so that the lightest particle carrying this quantum number could be a dark matter candidate.}, as long as the confinement scale, $\Lambda_3$ of the resulting $SU_3 (P)$ $N_F = N_C$ model was lower than $800-900$ GeV.  This scale arose from two {\it a priori} constraints:  the fixed line value of the couplings at $10^{16}$ GeV was barely perturbative (so that we could calculate the running) and the masses of the two heavy trianons are taken to be TeV scale, since they originate from the same diagrams that generate the SUSY breaking scale.

To avoid both light charged particles from the trianon sector, and too light a gluino, we choose the two heavier trianons to be colorless.  We will assume that the colored trianon Majorana mass is close to the scale $\Lambda_3$ .  The colored trianons are thus analogous to the strange quark in QCD, at the boundary of the region of quark masses where chiral perturbation theory works.  These masses fit between $\Lambda_3$ and the cutoff scale $4\pi \Lambda_3$.  By contrast, the colorless trianons should be thought of analogous to the charmed quark, or perhaps a slightly lighter quark, which is at the boundary where asymptotic freedom becomes a good approximation to heavy quark physics.   We will discuss issues of tuning in the conclusions.

\section{Gluino and squark masses}

In \cite{tbtj} the authors assumed that the gluino mass was Majorana, generated via loops of colored trianon fields as in standard direct gauge mediation models.
However, we failed to notice a larger contribution, coming from the following D term 

$$\delta {\cal L}_{gluino} = \int d^4 \theta\  (g(M, M^a ) D_{\alpha} M^a W_a^{\alpha} + c.c.) .$$  Our convention for operators appearing in D terms is that a function of complex variables is interpreted as a function of the variables and their complex conjugates.  We've also taken the non-perturbative scale $\Lambda_3$ of $SU_P (3)$ to be our unit of mass, and set it equal to $1$.   

Note that, just as hadron magnetic moments in QCD are proportional to the electromagnetic coupling $e$, this term is proportional to the TeV scale value of $g_3$, with no further loop suppression. Since $g_3 \sim 1.22$, this is a number of order $1$ in $\Lambda_3$ units.  If $M$ has a non-vanishing $F$ component, then this operator will give rise to a Dirac mass mixing the gluino field with the fermionic component of $M^a$.  

Another important operator has the form
$$ \int d^4 \theta\ h(M ) M_a M^a .$$  This gives rise to a Majorana mass for the fermionic component of $M^a$, if $M$ has a non-vanishing F component.  Since the Majorana and Dirac masses are of the same order, we will obtain two mass eigenstates whose masses differ only by a formal factor of $g_3^2$.   The strong interaction uncertainties in this calculation combine with the known value of $g_3$ to make the contributions comparable and we cannot really determine the field content of the lowest mass eigenstate.  Their splitting is $o[(F_M / \Lambda_3^{3/2})^2 ]$ .   By contrast, the Majorana mass term of the gluino field  is suppressed by a single QCD loop factor and is of order $ \frac{\alpha_3}{4\pi} F_M / \Lambda_3 $.   If $F_M$ and $\Lambda_3$ are comparable, this is smaller than the Dirac mass by a factor $\sim \frac{1}{150}$ .  The Pyramid scheme is thus one in which the Dirac mass of the gluino dominates its Majorana mass, without fine tuning.

It is worth recalling why we think that the standard contribution to the gluino mass is suppressed by $\frac{\alpha_3}{4\pi}$ in this model, while the operator that gives the gluino Dirac mass is $o(g_3)$, even though the messenger sector is strongly coupled.  If we calculate the two operators in $SU_P (3)$ perturbation theory, then all loops come with factors of the strong $SU_P (3)$ coupling for the first operator, while there is an uncompensated factor of $\frac{\alpha_3}{4\pi}$ in the two point functions that give rise to the gluino Majorana mass.  It is exactly analogous to the relation between the computation of hadron magnetic moments and hadronic contributions to electron-positron annihilation in QCD.  Apart from the perturbation theory argument, we see the factor of $\frac{\alpha}{4\pi}$ when we consider the contribution of a charged hadron anti-hadron pair to the current two point function.  

In Minkowski signature, the two point function is strongly enhanced on resonance if there are resonances with the quantum numbers of the current.  However, what is relevant for the computation of the effective action is the Euclidean two point function.
As shown in \cite{pqw}, in an asymptotically free theory, the Euclidean function gives an average over resonance behavior, and of course has the perturbative suppression.  
We are not calculating at a scale $\gg \Lambda_3$ but I see no reason to doubt the extra loop suppression in this regime.

Squared squark masses are generated by a {\it single} QCD loop diagram, with two insertions of the Dirac mass term of the gluino\cite{fnw}.  These diagrams are convergent.
The conventional logarithmically divergent term is proportional to the small gluino Majorana term, and is negligible since the logarithm is multiplied by $.01$ and the scale of the cutoff is low, since $F$ is so small.  The squark masses are flavor-blind, and of order $ \sqrt{\frac{\alpha_3}{4\pi}} m_{1/2}^{(3)}$.
The model thus predicts that squarks are ten times lighter than gluinos.  There is another, potentially important contribution to the squark mass matrix coming from the $F$ terms of the Higgs fields.  This will split the two stop masses from the other squarks and from each other.  One of the stops will be lighter than the first and second generation squarks.  We will discuss this in more detail when we examine the Higgs potential.

The LHC bounds on squarks, in the limit of large gluino mass, are between $.8$ and $.9$ TeV.  Our estimates then put the gluino at $8-9$ TeV, perhaps out of reach of the LHC.  One must be cautious though since there are strong interaction uncertainties in the ratio between gluino and squark masses.  It is not inconceivable that the gluino
could be a factor of $4$ lighter than our estimate.  

We can combine our estimates of the gluino mass from experimental bounds, with the CSB estimate of $F$, to obtain an estimate for $\Lambda_3$.  
$$8-9\ {\rm TeV} \sim \frac{900\ {\rm TeV}^4}{\Lambda_3^3} ,$$ so that $$\Lambda_3 \sim 4 - 5 {\rm TeV} .$$  There are major strong interaction uncertainties in this estimate, so it is not obviously inconsistent with the bound $\Lambda_3 < 900 GeV$ of \cite{bfk} .  Nonetheless, this tension might lead us eventually, to search for a different dark matter candidate in the Pyramid scheme.  It's possible for example that some components of the three singlets in the model could be dark matter candidates.   The model with all three pairs of trilinear couplings $(\lambda_p, \tilde{\lambda}_p)$ has a real fixed line and no upper bound on the value of $\Lambda_3$.  

The electroweak gauginos do not get a Dirac mass in this model.  The Majorana masses of the electroweak trianons are large enough compared to $\Lambda_3$ that the lightest states produced by the $SU(2) \times U(1)$ adjoint chiral bilinears in these fields are not degrees of freedom in the low energy Lagrangian below $4\pi \Lambda_3$.  We can think of them as having Majorana masses  $M_i > 4\pi \Lambda_3$.  The electroweak version of the operator which gives rise to the gluino Dirac mass, is much smaller and gives a seesaw contribution to the electroweak gaugino Majorana masses, of order
$$m_{1/2}^{(i)}\sim \frac{g_i^2 |F_M|^4 }{\Lambda_3^6 M_i} .$$   This is likely to dominate the usual gauge mediated contributions, which have an extra factor of $\frac{1}{16\pi^2} $.  

The idea that a Dirac mass for the gluino could lower the bounds on squark masses, and ameliorate the $\sim 1\% $ tuning of parameters required in generic supersymmetric models, has been pursued forcefully by G. Kribs and collaborators, as well as a number of other authors\cite{kribsetal} .  
These authors argue that the lower bounds on squark masses of the first two generations are reduced to about  $800 - 900$ GeV.  They explore other aspects of the phenomenology of Dirac gluino scenarios.

The authors of \cite{kribsetal} introduce a fundamental adjoint chiral multiplet\footnote{The exception is the last reference in this list, of which I was unaware at the time I wrote this paper.  These authors introduce composite adjoints, but their unification is in $SU(5)$ and the compositeness scale is much higher than that of the Pyramid scheme. I'd like to thank Jessica Goodman for bringing this paper to my attention.}.  To the extent that unification has been studied in this context, one assumes adjoints of the full unified group, to preserve one loop coupling unification.   Dirac masses for electroweak gauginos suppress the already small D-term contribution to the Higgs potential. In marked contrast, in the Pyramid Scheme the adjoint chiral multiplet has a compositeness scale of order $\Lambda_3$.  One loop coupling unification, and perturbative values of the standard model couplings up to the unification scale, is guaranteed by the underlying UV model, where the computation is done in terms of the trianons.  Indeed the Pyramid scheme was constructed as the only strongly coupled extension of the MSSM, which preserved coupling unification and was compatible with the very low scale of SUSY breaking implied by CSB.  There will be a threshold correction, which is hard to compute, to coupling renormalization through the strongly coupled regime.  This is of order the two loop correction to the beta functions.
Since any low energy model has uncontrolled threshold corrections at the unification scale, I have never seen any reason to worry about precision gauge coupling unification at the two loop level.  One should only do so if one has a model that is UV complete beyond the unification scale.  

The Pyramid Scheme version of Dirac/mixed gaugino masses thus seems to evade most of the objections of \cite{mina}

\section{The Higgs Potential and the mass of the lightest Higgs}

The Pyramid Scheme also has at least three standard model singlet fields, $S_i$.  It's worth recalling the reason for introducing these fields, since it does not follow the standard logic of effective field theory.  The basic assumption of CSB is that SUSY breaking is a consequence of interactions of the gravitino with quantum gravity degrees of freedom on the cosmological horizon.   Since SUSY is a local gauge symmetry in effective field theory, there {\it must} be an effective description of this exotic mechanism as spontaneous SUSY breaking.  The interactions with the horizon are incorporated as operators in effective field theory, which violate a discrete R symmetry that becomes exact for vanishing c.c..  These R violating terms do not obey the constraints of naturalness, and their UV running will be softened by the fact that the diagrams which generate them have internal gravitino lines that extend out to the horizon.

If one considers only the non-singlet sector of the Pyramid Scheme, there is no way to break SUSY.  The singlets are added as a minimal way to break SUSY, and it is achieved via the O' Raifeartaigh mechanism.  This requires\cite{nelsonseiberg} a choice of parameters in the superpotential that is not technically natural, but these parameters come from a special set of diagrams, with a single gravitino "loop", closed by an interaction with a featureless thermal bath on the horizon, so the usual field theory criteria do not apply.

The models were constructed so that the discrete R symmetry forbids all terms in the MSSM, through dimension $6$, which violate $B$ and $L$, apart from the dimension $5$ operator that generates neutrino masses.  It also forbids the $\mu H_u H_d$ term.  We found however that couplings $\sum \beta_i S_i H_u H_d$ were allowed by the R symmetry, as well as a coupling $\alpha_i S_i T_3 \tilde{T}_3 \rightarrow \Lambda_3 \sum \alpha_i S_i M$.  There is thus an increase in the Higgs quartic coupling, a ``tree level" potential that tends to destabilize the origin $H_u = H_d = 0$, and a mixing of the singlet with the Higgs.    These effects are increased by increasing $\sum |\beta_i |^2$.

Once one has chosen a superpotential that gives SUSY breaking, the computation of the effective potential is complicated by the fact that the Kahler potential of the fields, $M$ and two linear combinations of the $S_i$ (which couple to the heavy electroweak trianons) is a highly non-trivial function determined by the strong $SU_P (3)$ dynamics.  In \cite{tbtj} we made crude approximations to this function in order to estimate the spectrum a superpartners. We found a spectrum consistent with LHC data, with a few percent tuning of couplings.  However, in addition to the major error of omitting the gluino Dirac mass, we also left out the one loop contribution to the Coleman-Weinberg potential induced by Higgs-singlet mixing.  
Higgs-singlet mixing is also an important issue because the LHC has found the properties of the light Higgs to be close to those expected in the standard model.
Too much mixing between the singlet and the lightest Higgs could lead to deviations from that prediction.  

The interrelated questions of $SU(2) \times U(1)$ breaking, fine tuning of the Higgs potential, and Landau poles in the singlet Yukawa couplings to the Higgs, are complicated by the nontrivial Kahler potential $K (M, S_i , )$\footnote{Recall that the $S_i$ dependence comes from the coupling of $S_i$ to the heavy electroweak trianons. Generically, it will depend on a plane in the three dimensional $S_i$ space. The vectors $\alpha_i$ and $\beta_i$ will have components both in and out of this plane.}.   The potential contains terms of the form $$K^{i\bar{j}} (C + 3\alpha M + \beta H_u H_d )_i  (\bar{C} + 3\alpha^* \bar{M} + \beta^* \bar{H_u}\bar{ H_d} )_{\bar{j}} ,$$ which could drive a non-zero VEV for the Higgs fields.  We've arranged the cubic polynomial $C$ so that the F-terms in this expression can't vanish, but it's hard to determine where the real minimum lies without knowledge of the Kahler potential.  The three terms are linearly independent for all values of $S_i$, but need not be orthogonal in the Kahler metric.  If that is the case it will be advantageous to have a non-zero VEV for $H_u H_d$.  

If the Higgs VEV is nonzero, the tree level prediction of the Pyramid scheme is that the ratio of Higgs VEVS, $\tan \beta = 1$.  This will be corrected by the top supermultiplet Coleman-Weinberg potential, which favors larger $\tan\beta$ because the IR divergence of the top contribution makes the origin of $h_u$ unstable, but $\tan \beta$ is probably close to $1$, so the $F$ component of $H_u$ might contribute a substantial splitting of the top squarks from the universal squark mass\footnote{The latter comes primarily from a one loop diagram with the Dirac gluino in the loop.}, if the coefficients $\beta_i$ are large.  Half of the top squarks will be lighter, and half heavier than the squarks of the light generations, to which the LHC bounds apply.  

In the NMSSM, large $\beta_i$ is required in order to set the lightest Higgs mass at $125$ GeV.  However it also leads to strong mixing between the Higgs and the singlets. One cannot evaluate the masses of states in the Higgs sector without knowledge of the Kahler potential.  Large values of $\beta_i$ also raise the specter of large Coleman-Weinberg corrections to the scalar potential, and to the prospect of Landau poles below the unification scale.

The whole question of the Higgs sector is thus bound up with the non-perturbative Kahler potential.  The impatient reader should recall that if it were not for the non-trivial Kahler potential, SUSY breaking would decouple from all but the singlet part of the spectrum.   Thus, while it seems that the Pyramid scheme contains suggestions of how issues of tuning might be resolved, the actual calculations that would verify this are currently beyond our competence.  It is probably worth while to calculate all of these effects in the approximations used in \cite{tbtj} to assess whether any amelioration of the fine tuning can be achieved with our new insights.  That will involve fairly heavy numerical analysis, and is beyond the scope of the present paper.

\section{Conclusions}

The main new result of this paper is that the Pyramid Schemes predict that the gluino is strongly mixed with a composite octet chiral field, and that this mixing gives the dominant contribution to its mass.  The leading contribution to squark masses is ten times smaller than the gluino mass and is flavor blind and UV insensitive.  The much smaller contribution from the one QCD loop Majorana mass of the gluino does have a logarithmic enhancement, but the cutoff in this model is low, because the highest scale of SUSY breaking in the model is $\sim 5\ TeV$, so it is insignificant.  

The analogous effect for electro-weak gauginos leads to a seesaw Majorana mass, because the composite adjoints to which they are coupled have large Majorana masses.  This new contribution appears to dominate the conventional gauge mediated values of these mass parameters by a factor of $16\pi^2$, but strong interaction uncertainties and the absence of more than crude upper bounds on the electroweak trianon masses make this estimate less than reliable.

The most important question for the Pyramid schemes now is a careful re-evaluation of the scalar potential to address the related questions of the origin of $SU(2) \times U(1)$ breaking, the mass and properties of the lightest Higgs boson, the little hierarchy problem, and Landau poles in the Yukawa couplings of the top quark and that of the singlet fields to the Higgs .   This will require numerical work and will be addressed in a future publication.

\section{Acknowledgements} 

 I'd like to thank D. Shih, S. Thomas, W. Shepherd, J. Goodman, and L. M. Carpenter for discussions about and comments on the manuscript.  I'd also like to thank G. Kribs for discussions about the literature on Dirac gauginos. The research reported here was partially supported by the Department of Energy.

\end{document}